\documentclass{moriond}
\usepackage{siunitx}
\RequirePackage{xspace}

\def\Journal#1#2#3#4{{#1} {\bf #2}, #3 (#4)}

\def\NPB{{\em Nucl. Phys.} B}
\def\PLB{{\em Phys. Lett.}  B}

\def\PRD{{\em Phys. Rev.} D}

\def\be{\begin{equation}}
\def\ee{\end{equation}}
\def\bea{\begin{eqnarray}}
\def\eea{\end{eqnarray}}

\def\Vcb{\ensuremath{|V_{cb}|}\xspace}
\def\Vub{\ensuremath{|V_{ub}|}\xspace}

\def\Bbar{\kern 0.18em\overline{\kern -0.18em B}{}\xspace}

\def\BBbar{\ensuremath{B\Bbar}\xspace}
\def\Bzb{\ensuremath{\Bbar^0}\xspace}
\def\BzBzbar{\ensuremath{B^0 {\kern -0.16em \Bzb}}\xspace}
\def\Dbar{\kern 0.2em\overline{\kern -0.2em D}{}\xspace}

\def\Dzbar{\ensuremath{\Dbar^0}\xspace}
\def\Dstm{\ensuremath{D^{*-}}\xspace}

\def\BtoDstlnu{\ensuremath{B^0 \to \Dstm \, \ell^{+} \, \nu_\ell}\xspace}

\def\Btopilnu{\ensuremath{B^0 \to \pi^- \, \ell^+ \, \nu_\ell}\xspace}

\def\BtoXlnu{\ensuremath{B \to X \, \ell \, \bar{\nu}_\ell}\xspace}
\def\BtoXenu{\ensuremath{B \to X \, e \, \bar{\nu}_e}\xspace}
\def\BtoXmunu{\ensuremath{B \to X \, \mu \, \bar{\nu}_\mu}\xspace}
\def\BtoXclnu{\ensuremath{B \to X_c \, \ell \, \bar{\nu}_\ell}\xspace}
\def\BtoXulnu{\ensuremath{B \to X_u \, \ell \, \bar{\nu}_\ell}\xspace}
\def\cosBY{\ensuremath{\cos\theta_{BY}}\xspace}

\def\RXemu{\ensuremath{R(X_{e/\mu})}\xspace}

\newcommand{\Photo}{}

\begin{document}
\vspace*{4cm}
\title{Semileptonic $B$-meson decays including $b\to c$ anomalies at Belle~II}

\author{H.~Junkerkalefeld on behalf of the Belle~II Collaboration}

\address{Physikalisches Institut der Rheinischen Friedrich-Wilhelms-Universität Bonn, 53115 Bonn, Germany}

\maketitle\abstracts{
We present recent measurements of semileptonic $B$-meson decays using a data sample collected at the $\Upsilon(4S)$ resonance by the Belle~II experiment corresponding to an integrated luminosity of \SI{189}{\per\femto\barn}. We determine the CKM-matrix elements \Vcb using untagged \BtoDstlnu decays and \Vub using untagged \Btopilnu decays. We test light lepton universality in two new analyses using hadronically tagged events. We report the first measurement of a complete set of five different angular distributions using \BtoDstlnu decays and we measure the branching fraction ratio of inclusive \BtoXenu and \BtoXmunu decays.}

\section{Introduction}

This work presents four recent measurements with semileptonic $B$-meson decays. All measurements use a data set corresponding to \SI{189}{\per\femto\barn} of $e^+e^-$ collisions at the $\Upsilon(4S)$ resonance and \SI{18}{\per\femto\barn} of collisions \SI{60}{\MeV} below the $\Upsilon(4S)$ resonance collected by the Belle~II experiment between 2019 and 2021. The Belle~II detector consists of several nested subsystems arranged around the interaction region that are described in more detail in Ref.~\cite{belle2_1,belle2_2}. Charge conjugation is implied in all physical processes and natural units are used.

\section{CKM matrix element determinations and branching fraction measurements}

The Cabibbo-Kobayashi-Maskawa (CKM) matrix relates the weak-interaction and mass eigenstates in the Standard Model (SM) of particle physics. Its values are fundamental parameters of the SM and have to be determined experimentally. A longstanding discrepancy persists between the two different determination methods to extract \Vcb and \Vub from semileptonic $B$-meson decays \cite{hflav}: the exclusive determinations, in which the hadronic part of the decay is explicitly reconstructed and inclusive approaches that are sensitive to all hadronic final state at the cost of experimentally more challenging distributions. This inconsistency currently limits the predictive power of precision flavor physics so that its understanding is crucial for future measurements.\par

\subsection{\Vcb determination using untagged \BtoDstlnu decays}\label{sec:untaggedBtoDstlnu}

In this analysis, we reconstruct \BtoDstlnu decays ($\ell = e, \mu$) with the subsequent decays $\Dstm \to \Dzbar \, \pi^-$ and $\Dzbar \to K^+\pi^-$. Light leptons are identified using particle identification likelihood ratios based on the different subdetectors. Neutral \Dzbar meson candidates are reconstructed by combining charged pion and kaon candidates with appropriate invariant masses. Subsequently, \Dstm candidates are formed by combining \Dzbar candidates and slow pion candidates.\par
The partner $B$ meson produced in the $\Upsilon(4S)$ decay is not explicitly reconstructed so that the signal $B$ meson's momentum direction has to be estimated differently. This is done in a novel approach based on the  angle \cosBY between the signal $B$ meson and the $Y = D^* + \ell$ system. Assuming correct reconstruction, \cosBY can be inferred from beam energy information and the reconstructed particle kinematics. A weighted average of possible momentum directions is calculated by combining the following additional information: the expected angular distribution of $B$ mesons along the beam axis due to the $\Upsilon(4S)$ polarization and additional rest-of-event (ROE) information of tracks and clusters not associated with the $Y$ system.\par
The directional information is used to derive the recoil parameter $w = (m_B^2+m_{D^*}^2-q^2)/(2 m_B m_{D^*})$ and the helicity angles $\cos\theta_\ell$, $\cos\theta_V$, and $\chi$, that are defined to be the angles between lepton and $W$ boson directions, $D$ and $D^*$ meson directions and the angle between the two decay planes. The four quantities are divided into 10 (8) equidistant bins each in which the background and signal yields are derived individually in a two-dimensional maximum-likelihood fit to \cosBY and the mass difference of the \Dstm and \Dzbar mesons $\Delta M$. Subsequently, effects of the finite reconstruction resolution, efficiency and detector acceptance are corrected.\par
By summing the partial decay rates of all kinematic variables and by averaging over $w$ and the helicity angles we obtain the branching fraction $\mathcal{B}(\BtoDstlnu) = (4.94 \pm 0.02^\text{stat.} \pm 0.22^\text{syst.})\,\text{\%}$ with excellent agreement between the electron and muon channel. Systematic uncertainties are dominated by slow pion reconstruction efficiency uncertainty and the external input $f_\pm/f_{00}$, i.e., the ratio of $\Upsilon(4S)\to B^+B^-$ to $\BzBzbar$ events. The CKM matrix element \Vcb and the form factor parameters of the BGL parametrization  \cite{BGL1,BGL2} are derived in a fit to the normalized decay rates using lattice QCD input for the normalization at zero recoil \cite{fnalmilc}. We find
\begin{equation}
	\Vcb = (40.9 \pm 0.3^\text{stat.} \pm 1.0^\text{syst.} \pm 0.6^\text{theo.}) \times 10^{-3} \, \text{.}
\end{equation}
The result lies between HLFAV's exclusive and inclusive world averages \cite{hflav}.
Additionally, lepton universality is tested in this analysis by comparing electron and muon angular properties as depicted in Figure \ref{fig:afb_results} and motivated in Section \ref{sec:afb}.

\subsection{\Vub determination using untagged \Btopilnu decays}

The experimentally and theoretically golden mode to extract \Vub exclusively is the untagged \Btopilnu channel in which the partner $B$ meson is not explicitly reconstructed. We present a new study of this decay \cite{pilnu}. Here, lepton candidates are identified via likelihood ratios. Signal pion candidates are required to be oppositely charged to the lepton candidate, to satisfy loose likelihood ratio requirements and tight track quality constraints.\par
The main challenge of this analysis is to suppress wrongly reconstructed \BBbar background events, that are subsequently divided into \BtoXclnu, \BtoXulnu and other \BBbar events, as well as continuum backgrounds mainly from $e^+e^- \to q\bar{q}$ ($q = u, d, s, c$) and $ee\tau\tau$ events. For this purpose, several boosted decision trees (BDTs) are trained to differentiate between signal and each background category separately in each kinematic region. The continuum suppression BDTs mainly use event-topology quantities and the \BBbar background suppression BDTs utilize quantities derived from the signal lepton and pion candidates and the ROE.\par
The signal and background yields are extracted in a binned two-dimensional likelihood fit to $\Delta E = E_B^\text{c.m.} - E_\text{beam}^\text{c.m.}$ and $M_{bc} = \sqrt{(E_\text{beam}^\text{c.m.})^2 - |\vec{p}_B^\text{\,\,c.m.}|^2}$ in six bins of $q^2 = (p_B - p_\pi)^2$ simultaneously. For this purpose, the direction of the signal $B$ meson is estimated in the approach introduced in Section \ref{sec:untaggedBtoDstlnu}.
We find a branching fraction averaged over leptons flavors of $\mathcal{B}(\Btopilnu) = (1.426 \pm 0.056^\text{stat.} \pm 0.125^\text{syst.})\times 10^{-4}$, which is consistent with the current world average. The result is limited by the size of the off-resonance data set that is used to constrain continuum modeling.\par
A $\chi^2$ fit is performed on the measured $q^2$ spectra to extract \Vub. Lattice QCD constraints on the eight BCL parameters from Ref.~\cite{BCL} are included as nuisance parameters and the lepton flavor averaged result is
\begin{equation}
	\Vub = (3.55 \pm 0.12^\text{stat.} \pm 0.13^\text{syst.} \pm 0.17^\text{theo.})\times 10^{-3}\;\text{.}
\end{equation}
The result is consistent with the current value obtained by HFLAV \cite{hflav} and the experimental uncertainties are limited by uncertainties on the modeling of continuum and $B\to\rho\ell\nu$ events.

\section{Novel tests of light-lepton universality}

In the SM all leptons share the same electroweak coupling, a symmetry referred to as lepton universality (LU). Recently, however, a $4\,\sigma$ evidence for lepton-universality violation (LUV) has been reported in a reinterpretation of Belle data in angular distributions of $B\to D^* \ell \bar{\nu}_\ell$ decays with light leptons \cite{bobeth}. Additionally, the branching-fraction ratio of semileptonic decay rates to $\tau$ leptons relative to the light leptons has been observed to indicate LUV in several experiments \cite{hflav}.\par

\subsection{Angular asymmetries of hadronically tagged \BtoDstlnu decays}\label{sec:afb}

We present the first dedicated light-lepton universality test using a complete set of angular asymmetry observables, called $A_\text{FB}$, $S_3$, $S_5$, $S_7$, and $S_9$. They are designed to be maximally sensitive to LUV and cancel most theoretical and experimental uncertainties \cite{angular_asymm1,angular_asymm2}. They are derived by constructing disjoint one- or two-dimensional integrals of the differential decay rate as a function of the recoil parameter $w$ and the helicity angles $\cos\theta_\ell$, $\cos\theta_V$, and $\chi^2$ as introduced in Section \ref{sec:untaggedBtoDstlnu}. They can be expressed as
\begin{equation}
	\mathcal{A}_x(w) = \left( \frac{\text{d}\Gamma}{\text{d}w}\right)^{-1} \left[ \int_0^1 - \int_{-1}^0 \right]\text{d}x \frac{\text{d}^2\Gamma}{\text{d}w\text{d}x} \; \text{,}
\end{equation}
with $x = \cos\theta_\ell$ for $A_\text{FB}$, $\cos 2\chi$ for $S_3$, $\cos\chi\cos\theta_V$ for $S_5$, $\sin\chi\cos\theta_V$ for $S_7$ and $\sin 2\chi$ for $S_9$. Lepton universality is tested by comparing the angular asymmetries of electrons and muons $\Delta \mathcal{A}_x (w) = \mathcal{A}_x^\mu(w) - \mathcal{A}_x^e(w)$. The integral is measured in three $w$ ranges: the full phase space ($w_\text{incl.}$), $w \in [1.0, 1.275]$ ($w_\text{low}$) and $w \in [1.275, \approx 1.5]$ ($w_\text{high}$).\par
One $B$ meson, the $B_\text{tag}$, is reconstructed in a fully hadronic decay using the Full Event Interpretation tagging algorithm \cite{FEI}. Only one $B_\text{tag}$ candidate is kept based on the highest value of the algorithm's output classifier. In events with such a candidate, we reconstruct \BtoDstlnu candidates with $\Dstm \to \Dzbar \pi^-$ and $\Dzbar \to K^+\pi^-$, $K^+\pi^-\pi^+\pi^-$, $K^+\pi^-\pi^0$, $K^+\pi^-\pi^+\pi^-\pi^0$, $K_\text{S}^0\pi^+\pi^-$, $K_\text{S}^0\pi^+\pi^-\pi^0$, $K_\text{S}^0\pi^0$, or $K^+K^-$. All tracks are required to originate from the $e^+e^-$ interaction point. Lepton candidates are identified based on likelihood ratios.\par
The signal and background yields are extracted in binned maximum-likelihood fits to distributions of $M_\text{miss}^2 = (p_{\Upsilon(4S)} - p_{B_\text{tag}}- p_{D^*} - p_\ell)^2$ which peak near zero for correctly reconstructed signal events and are more broadly distributed for backgrounds. The fitted yields are corrected for selection and detector acceptance losses and bin migrations. Statistical uncertainties largely dominate and are generally one order of magnitude larger than uncertainties due to bin migrations or systematic uncertainties.\par

\begin{figure}[htbp]
	\centerline{\includegraphics[width=\linewidth]{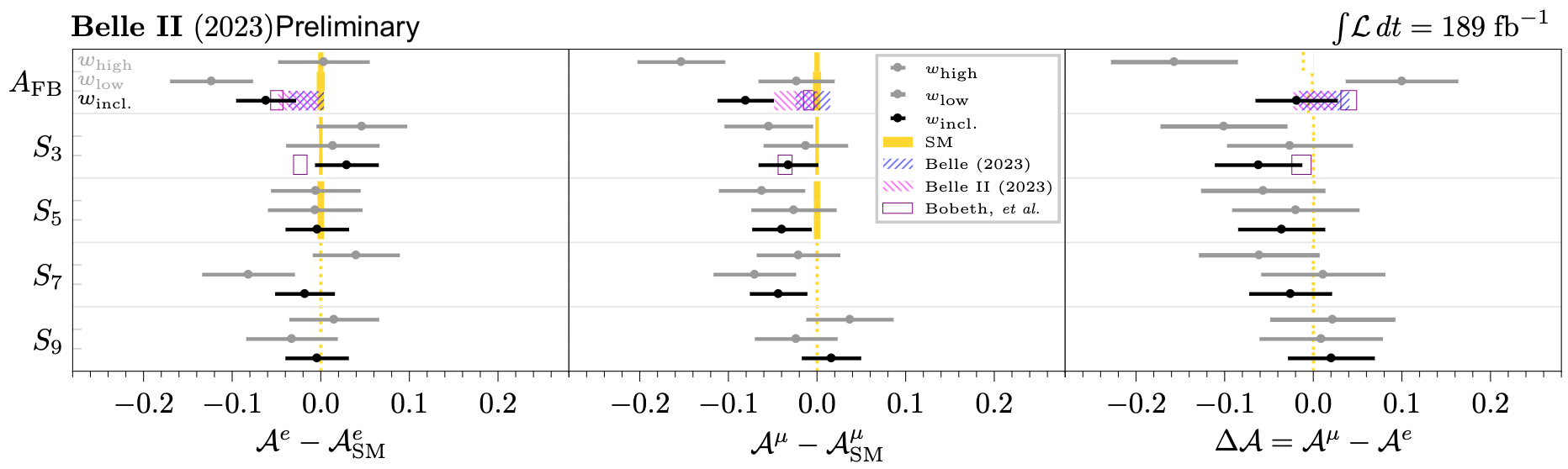}}
	\caption[]{The measured asymmetries and asymmetry differences are depicted and compared to recent Belle \cite{mprim} and Belle~II (cf. Section \ref{sec:untaggedBtoDstlnu}) results, calculations from Bobeth \textit{et al.} \cite{bobeth}, and SM expectations.}
	\label{fig:afb_results}
\end{figure}

The results are summarized in Figure \ref{fig:afb_results}. In $\chi^2$ tests in each of the three $w$ regions, the measured angular asymmetries are compatible with the Standard-Model expectation with a minimum $p$ value of $0.12$ and no indication for LUV is observed.

\subsection{Branching-fraction ratio measurement of inclusive \BtoXlnu decays}

In this analysis \cite{RXemu}, we report the first measurement of the inclusive ratio of branching fractions $\RXemu = \mathcal{B}(\BtoXenu) / \mathcal{B}(\BtoXmunu)$. One $B$ meson is reconstructed fully hadronically. On the signal side, the lepton candidate is demanded to have a momentum in the rest frame of the signal $B$ meson of $p_\ell^B > \SI{1.3}{\GeV}$ to suppress hadrons faking leptons, leptons emerging from secondary cascade decays and lepton daughters of $B\to X \tau \nu$ decays. Muon candidates are identified using likelihood ratios and electron candidates are selected using a multiclass BDT  \cite{eBDT}.\par
Continuum backgrounds are suppressed using a BDT trained on event-shape quantities. The remaining continuum events are described using the off-resonance data sample.\par
The signal yields are extracted in a binned maximum-likelihood template fit to the $p_e^B$ and $p_\mu^B$ spectra in 10 equally sized bins per lepton flavor simultaneously to account for correlated systematic uncertainties between both flavors. Continuum background yields are constrained using off-resonance data. \BBbar backgrounds get a Gaussian constraint from a fit to data in a background-enriched same-charge control channel containing events with two $B$ mesons reconstructed with the same flavor. Statistical and systematic uncertainties are included in the fit as nuisance parameters.\par
We find and \RXemu value of 
\begin{equation}
	\RXemu = 1.033 \pm 0.010^\text{stat.} \pm 0.019^\text{syst.} \; \text{.}
\end{equation}
The systematic uncertainties are dominated by uncertainties associated to the lepton-identification efficiency and hadron-misidentification weights. Branching-fraction and form-factor uncertainties of resonant and non-resonant \BtoXlnu decays mostly cancel in the ratio and hardly contribute. We find the result to be stable against the lower threshold of $p_\ell^B$ by changing it from the nominal value of \SI{1.3}{\GeV} to $1.1$, $1.2$ and \SI{1.4}{\GeV}. Our result is consistent with the Standard Model prediction \cite{kvos} within $1.2\,\sigma$ and it is the most precise branching-fraction based test of electron-muon universality in semileptonic $B$ decays to date.

\end{document}